\documentclass[english]{article}
\usepackage{float}
\usepackage{amsmath}
\usepackage{graphicx}
\usepackage{babel}

\usepackage[obeyspaces]{url}
\usepackage{hyperref}
\usepackage{setspace}

\usepackage{authblk}

\usepackage{caption}
\begin{document}

\title{Time, Space and Social Interactions: Exit Mechanisms for the Covid-19 Epidemics}

\author[1,2]{Antonio Scala}
\author[3]{Andrea Flori}
\author[4,5]{Alessandro Spelta} 
\author[1]{Emanuele Brugnoli} 
\author[1]{Matteo Cinelli} 
\author[6,1]{Walter Quattrociocchi}
\author[3,5]{Fabio Pammolli}

\affil[1]{Applico Lab, CNR-ISC}
\affil[2]{Big Data in Health Society}
\affil[3]{Impact, Department of Management, Economics and Industrial Engineering, Politecnico di Milano}
\affil[4]{Univ. di Pavia}
\affil[5]{Center for Analysis Decisions and Society, Human Technopole and Politecnico di Milano}
\affil[6]{Univ. di Venezia 'Ca Foscari}

\maketitle

\begin{abstract}
\onehalfspacing
We develop a minimalist compartmental model to study the impact of mobility restrictions in Italy during the Covid-19 outbreak. We show that an early lockdown shifts the epidemic in time, while that beyond a critical value of the lockdown strength, the epidemic tend to restart after lifting the restrictions. As a consequence, specific mitigation strategies must be introduced. We characterize the relative importance of different broad strategies by accounting for two fundamental sources of heterogeneity, i.e. geography and demography. First, we consider Italian regions as separate administrative entities, in which social interactions between age classs occur. Due to the sparsity of the inter-regional mobility matrix, once started the epidemics tend to develop independently across areas, justifying the adoption of solutions specific to individual regions or to clusters of regions. Second, we show that social contacts between age classes play a fundamental role and that measures which take into account the age structure of the population can provide a significant contribution to mitigate the rebound effects. Our model is general, and while it does not analyze specific mitigation strategies, it highlights the relevance of some key parameters on non-pharmaceutical mitigation mechanisms for the epidemics.

\end{abstract}

\section{Introduction}
Different epidemic models and approaches contribute to identify specific mechanisms relevant for policy design \cite{keeling2005models}.
At present, although the World Health Organization (WHO) organizes regular calls for Covid-19 modelers to compare strategies and outcomes, policymakers barely handle the discrepancies between the proposed models\footnote{See, e.g.: https://www.sciencemag.org/news/2020/03/mathematics-life-and-death-how-disease-models-shape-national-shutdowns-and-other}. 

To contain the Covid-19 epidemic, governments worldwide have adopted severe social distancing policies, ranging from partial to total population lockdown \cite{di2020timing}. Restrictions have led to a sudden stop of economic activities in many sectors, while the majority of Covid-19 infections affect active population (i.e. the age class between 15–64 years) \cite{surveillances2020epidemiological}. Overall, the impact of contagion and lockdown measures on health and on economic activities is substantial and pervasive. 

Against this background, we introduce a model-based scenario analysis for Covid-19, and we highlight how geographical and demographic variables influence the epidemic spreading and the effects of lockdown solutions, while providing some general indications on relevant exit mechanisms  \cite{anderson2020will}. 

The general behavior of our framework holds for the vast class of epidemic models where transmission rate is proportional to the number of susceptible people times the density of infected. We focus on the determinants of short-term interventions in response to an emerging epidemic when geographic and demographic compartments are included in the model. Our goal is general in nature, since we focus on two relevant decomposability conditions, under which partial dynamics influence the overall configuration of the system (see, e.g., \cite{simon1961aggregation,ando1963near,simon1996architecture,courtois2014decomposability}). We study how a) mobility restriction measures and b) timing of the lockdown lift affect the total fraction of infected, the peak prevalence, and, possibly, the delay of the epidemic. Our analysis identifies two fundamental sources of heterogeneity in the diffusion process: regional boundaries and age classs \cite{anderson2020will}. We show how such dimensions can shape policy interventions aiming at containing the epidemic, irrespective of any detailed quantitative predictions on specific micro level measures. 

This paper aims to contribute to the extant literature on trade-offs between mitigation, i.e. slowing down the epidemic contagion, and suppression, i.e. temporarily compressing the risk of contagion \cite{anderson2020will,ferguson2020report,prem2020effect}. Notwithstanding micro data on individual profiles are not taken into account, our simple compartmental model based on geographical and age classes uncovers relevant aspects, which provide some guidance to policy makers. First, we show that an early lockdown shifts the epidemic in time and that the delay is proportional to the anticipation time, with an intensity which grows with the strength of the lockdown. Beyond a critical threshold, the epidemic would tend to fully recover its strength as soon as the lockdown is lifted. 
As a consequence, specific mitigation strategies must be prepared during the lockdown. To provide some guidance on the relative importance of different general strategies, we first study how the sparsity of the matrix representing mobility flows across administrative regions influences the observed delays of the contagion. The relative strength of infra regional mobility with respect to inter regional mobility flows implies that, once the epidemic has started, it then tends to develop independently within each region \cite{chinazzi2020effect}. Second, we study the impact of patterns of interaction within and between age classs, and we find its structure to be of primarily importance to estimate post-lockdown effects. According to our results, age-based mitigation strategies can be a key ingredient to contain rebound effects.

\section{Model}

To analyze mobility-restriction policies, we introduce a minimalist compartmental model \cite{prem2020effect,Bailey1975book}. Although many models, both mechanistic, statistic and stochastic \cite{liu2020reproductive}, have been proposed for the Covid-19 infection, data collected from national healthcare systems suffer from the lack of homogeneous procedures in medical testing, sampling and data collection \cite{casella2020can}. Not to mention the difficulties in assessing the impact of variability in social habits during the epidemics \cite{prem2020effect,funk2010review}.  
Moreover, especially in the early phases of the epidemic -- i.e. the ones characterised by an exponential growth -- different models sharing a given reproduction number $R_0$ can fit the data with equivalent accuracy (see discussion in the APPENDIX about fitting initial parameters). 
For these reasons, our aim is to focus on some fundamental qualitative scenarios and not on detailed predictions. We adapt the $SIR$ model, the most basic epidemic model for flu-like epidemics, to the observed data available in the Italian case.  

The model relies on four compartments, namely: $S,I,O,R$. Hence, $S$(usceptible) individuals can become $I$(nfective) when meeting another infective individual, $I$(nfectives) either become $O$(bserved) -- i.e. present symptoms acute enough to be detected from the national health-care system -- or are $R$(emoved) from the infection cycle by having recovered; also $O$(bserved) individuals are eventually $R$(emoved) from the infection cycle\footnote{We are implicitly absorbing the number of deaths in the $R$(emoved) compartment of the model, that therefore comprises both the $R$ecovered people (who hopefully have developed antibodies and are not anymore susceptible) and the small fraction of those who did not overcome the epidemics}. In the case of Covid-19, it is not clear yet if there is an asymptomatic phase \cite{bai2020presumed,nishiura2020estimation}; in the model, we implicitly assume that asymptomatics are infective and their removal time is the same of the $I$ class. The model is described by the following differential equations:
\begin{equation}
\begin{split}
\partial_t S &= - \beta S\,\frac{I}{N} \\
\partial_t I &= \beta S\,\frac{I}{N} - \gamma I \\
\partial_t O &= \rho \gamma I - h O \\
\partial_t R &= (1-\rho) \gamma I + h O
\end{split}
\label{eq:SIOR}
\end{equation}
 $N=S+I+O+R$ is the total number of individuals in a population, the transmission coefficient $\beta$ is the rate at which a susceptible becomes infected upon meeting an infected individual, $\gamma$ is the rate at which an infected either becomes observable or is removed from the infection cycle. Like the $SIR$ model, the basic reproduction number is $R_0=\beta/\gamma$; the extra parameters of the $SIOR$ model are $\rho$, the fraction of infected that become observed from the national health-care system, and $h$, the rate at which observed individuals are removed from the infection cycle. Notice that we consider that $O$(bserved) individuals not infecting others, being in a strict quarantine.

\section{The Italian Lockdown}

The Italian lockdown measures of the $8^{\text{th}}$ and $9^{\text{th}}$ of March \cite{dpcm8marzo,dpcm9marzo} aimed to change mobility patterns and to reduce the intensity of social contacts, through quarantine measures and to an increased awareness of the importance of social distancing.
We analyze an extensive data set on Facebook mobility data\footnote{Those data are part of the Facebook project ``Data for Good'', and illustrate mobility patterns of fb users, who allowed the social network to track their location. See
https://dataforgood.fb.com/docs/Covid-19/} \cite{buckee2020aggregated}; our analysis confirms that the lockdown has reduced both the travelled distance and the flow of travelling people.

We consider the effects of lockdown measures on the parameters of our model. Lockdown is a non-pharmaceutical measure; hence the rate $\gamma$ is the most unaffected parameter, since it is related to the ``medical'' evolution of the disease. Analogous arguments apply to the rate $h$ of exiting a condition serious enough to be observed and to the probability $\rho$ of being observed by the national healthcare system  (although $\rho$  could be influenced variations in testing schemes and alert thresholds). On the other hand, the transmission coefficient $\beta$ can be thought as the product $C\lambda$ of a contact rate $C$ times a disease-dependent transmission probability $\lambda$. Hence, if we assume that the speed of Covid-19 mutation is irrelevant on our timescales, lockdown strategies mostly influence $\beta$ by reducing the contact rate $C$ between individuals.

To adapt the $SIOR$'s parameters to the Italian data \cite{iss_epicentro}, we compare the reported cumulative number of Covid-19 cases $Y^{\text{Obs}}$ with the analogous quantity $Y^{\text{model}}=\int \rho \gamma Idt $ in our model. We want to stress that our model fitting is not aimed to produce an accurate model for detailed predictions, but to work in a realistic region of the parameter space.

We first estimate model's parameters by least square fitting on the pre-lockdown period. Since in such range the data  $Y^{\text{Obs}}$ show an exponential growth trend, we are possibly observing a very early phase of the epidemic, where $\beta -\gamma$ equals the growth rate of $Y^{\text{Obs}}$ (see APPENDIX for observations on the choices of initial parameters). For fixed $\beta -\gamma$, the time of the epidemic start (that we conventionally assume as the time $t_0$ where the number of infected is $1$) and the fraction $\rho$ of serious cases observed by the national healthcare service, allows to vary the values of $\beta$ and $\gamma$ as long as their difference is fixed. Hence, estimating \emph{medical} parameters as the rate $\gamma$ of escaping the infected state is paramount for calibrating mathematical models.

In response to the outbreak of Covid-19, several estimates of model parameters have been proposed in the literature, revealing a certain amount of uncertainty about some fundamental variables of the epidemic contagion. The European Centre for Disease controls reports an infection time duration $\tau_I$ between $5$ and $14$ days \cite{ECDC8apr2020report}; in our model, we will use $\tau_I=10$ (i.e. $\gamma=\tau_I^{-1}=1/10\,days^{-1}$). According to a report of ISS, the Italian National Health Institute, the time from the start of serious symptoms (i.e. when one gets ``observed'' from ISS) to the resolution of the symptoms can be estimated as $\tau_H \sim 9\,days$ \cite{ISS30march2020report}, corresponding in our model to a value $h=1/9\,days^{-1}$. Notice the analysis of 12 different models \cite{liu2020reproductive} reports varying estimates for the basic reproduction number $R_0$, ranging from $1.5$ to $6.47$, with mean $3.28$ and a median of $2.79$. 

From fitting the $15$ days of $Y^{obs}$ (pre-lockdown phase) and by performing a bootstrap sensitivity analysis of the parameters, we obtain $\beta-\gamma \sim 0.25 \pm 0.01$ 
and $t_0=-30 \pm 5\,days$ 
by assuming that $\rho=40\%$.  Varying $\rho$ in $[10\%,100\%]$ varies $\beta-\gamma$ in $[0.22,0.27]$. On the other hand, for fixed $\beta-\gamma$, $R_0$ would vary linearly with $\tau_I$; as an example, $R_0$ varies in $[2.5,4.5]$ for the literature parameters $\tau_I\in[5, 14]$; accordingly, to adjust the difference in growth rate, $t_0$ varies in $[26,32]$. However, despite the variability of the parameter range, the qualitative behavior of the model -- and hence our analysis of the key factors of the epidemic evolution -- is unchanged.

We then assume that, after the lockdown day $t_{\text{Lock}} =15$ (corresponding to the $9^{th}$ of march), contact rate drops down by a factor $\alpha$ and hence $\beta\to\alpha\beta$. By fitting the observed data $Y^{obs}$ for a symmetric period of $15$ days after $t_{\text{Lock}}$, and by performing a bootstrap sensitivity analysis, we find $\alpha=0.49\pm0.01$, i.e. a $\sim 50\%$ reduction in infectivity and hence in $R_0$. Our figure is in line with the observed reduction in $R_0$ in response to the combined non-pharmaceutical interventions, that across several countries has and average reduction of $64\%$ compared to the pre-intervention values \cite{flaxman2020report}. Notice that Facebook  mobility data show a post-lockdown reduction in mobility of $15\%$ at regional level and of $73\%$ at inter-regional level; however, as we will point out later, mobility has a strong impact at the beginning of the epidemics in new regions/countries, while it has much lower effects on the evolution of the epidemics in a region/country. 

In the following, we will use the parameters of Tab.~\ref{tab:Params}, corresponding to a basic reproduction number $R_0=3.5$. Moreover, since patients in intensive care represent the highest burden for health facilities, in the graphs of the paper we will indicate the number of patients in intensive care, estimated as $3.5\%$ of the total patients by using the figures reported by ISS \cite{iss_epicentro}.  

\begin{table}
    \centering
    \begin{tabular}{|c|c|c|}
    \hline
$\beta = 0.35\,day^{-1}$ & $\gamma=10^{-1}\,day^{-1}$ & $h=1/9\,day^{-1}$ \\ 
\hline
$t_0=-30\,days$ & $\rho=40\%$ & $\alpha=0.49$  \\
\hline
    \end{tabular}
    \caption{Standard parameters used for the $SIOR$ model in the paper.}
    \label{tab:Params}
\end{table}

\section{National scenarios and exit mechanisms}

Since we are interested on the factors driving the exit dynamics from lockdown, and not on the detailed analysis of realistic scenarios, we consider several lockdown scenarios, where the lockdown is abruptly lifted and the system let return to the pre-lockdown parameters. Such an approach clearly describes a worst-case estimate of the intensity of the second wave of the contagion after the conclusion of the lockdown. Hence, we consider several simplified scenarios, where we use the $SIOR$ model described by System~\ref{eq:SIOR} with the parameters of Tab.~\ref{tab:Params}. First, in the simple case of a $SIOR$ model fitted on Italian data, we analyze how the post-lockdown dynamics changes according to different starting dates of the epidemics and to different levels of the restrictions implemented by the national authorities. Then, we study the effect of explicitly considering Italy as a collection of separate administrative entities (Regions); finally, we consider the effects of social interactions across age class. 

\begin{figure}
    \centering
    \includegraphics[width=\columnwidth]{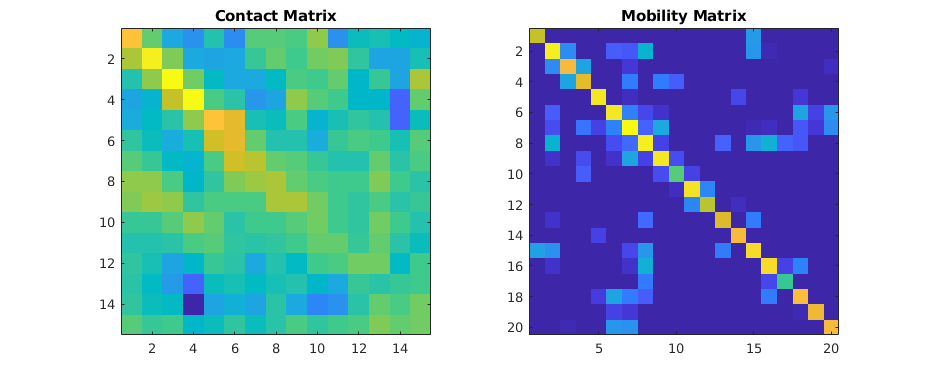}
    \caption{Left Panel: social contact matrix, from \cite{mossong2008POLYMOD}. Right panel: inter-regional mobility matrix, from the Facebook project ``Data for Good''. 
    The intensity of a color maps the strength of a matrix element (light colors: high values; dark colors: low values). The inter-age social mixing matrix is dense; hence age classes dynamics are strongly coupled. The inter-regional mobility flows is very sparse (i.e. off diagonal elements are order of magnitudes lower than diagonal elements): this mean that most of the people travel within the same region of origin; hence, the regional dynamics can be considered ``almost'' decoupled.}
    \label{fig:contactsVSmobility}
\end{figure}

Interestingly, mobility flows \cite{buckee2020aggregated} and inter-age social mixing \cite{mossong2008POLYMOD} lie at the two opposite range of modelling.  In fact, the regional social contact matrix is dense (Fig.~\ref{fig:contactsVSmobility}, left panel), indicating that age classes dynamics are strongly coupled. On the other hand, the inter-regional mobility matrix is very sparse (Fig.~\ref{fig:contactsVSmobility}, right panel), indicating that regions have their own independent dynamics. 

We first consider a simple exit strategy consisting in lifting the lockdown at a time $t_{\text{Unlock}}$ after the peak of $O$ has occurred. For instance, we hypothesize that infection proceeds uncontrolled up to time  $t_{\text{Lock}}$; in the following lockdown period $[t_{\text{Lock}},t_{\text{Unlock}}]$, the transmission coefficient $\beta$ is reduced by a factor $\alpha$; finally, $\beta$ returns to its initial value and herd immunity is responsible for the dampening of the epidemics.

\begin{figure}
\begin{centering}
\includegraphics[width=0.7\columnwidth]{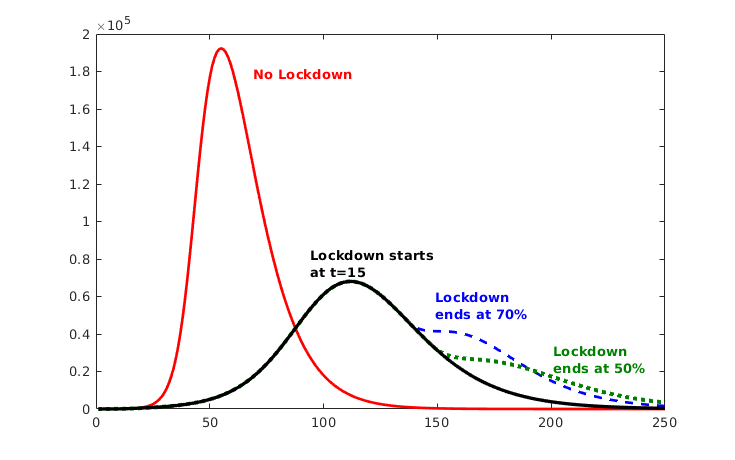}
\par\end{centering}
\caption{Comparison of the scenarios where the lockdown is relaxed after the percentage of people with visible symptoms ($O$) is reached the $70\%$ and the $50\%$ of the reported cases peak. Lifting the lockdown earlier has the epidemics disappear faster, but has higher impact on the number of hospitalized and intensive care patients; moreover, lifting the lockdown too early can result in a rebound of the number of cases.}
\label{fig:Scenario1}
\end{figure}

Our results show that the lockdown lowers the peak of $O$ - i.e. the  individuals with noticeable symptoms - to $\sim 70 \%$ of the free epidemic one, but it also doubles the time of its occurrence from $\sim 1.9$ months to $\sim 3.8$ months: an extremely obnoxious effect for the sustainability conditions of the economy of a country. However, since the number of hospitalized patients and - most importantly - the number of patients in intensive care is only a fraction of $O$, lowering the peak puts less stress on the healthcare system. The ideal situation would be to have accurate data, an accurate model and accurate estimates of the parameters; as an example, in our model lifting the lockdown when the number of infected people per unit time $\beta S(t)I(t)/N$ is lower that the average number of recovering people $\gamma I(t)$ would ensure that the number of infections would continue to decrease. In real life, situations are more fuzzy: having not enough information, we could decide to resort on some heuristics, like lifting the lockdown after the observed people $O$ have dropped to a suitable percentage of the maximum peak. As an example, after $\sim 4.7$ months the peak has reduced to $70\%$ of its initial value, while after $\sim 5.2$ months to $50\%$, i.e. $\sim 0.5$ months later. Notice that, the earlier the lockdown is lifted, the faster $O$ decays to zero even if it starts from higher figures and could even experience a rebound. All such effects are shown in Fig.~\ref{fig:Scenario1}.

Our framework sustains the identification of several mechanisms. The first is related to the timeliness of the lockdown, i.e. to the choice of anticipating $t_{\text{Lock}}$. As expected, early lockdown (i.e. well before the ``free'' infection peak) reduces the height of the peak without much moving it forward in time. Conversely, lifting the lockdown too soon can make epidemic start again and reach values higher than the ones before the release. A peculiar and counter-intuitive effect can be generated if the lockdown is anticipated: in fact, a too early lockdown delays the start of the epidemic without attenuating its severity (see APPENDIX for the description of the effects of varying lockdown time). In other words, an early lockdown ``buys'' time, but it postpones the problem without mitigating its severity.

Another effect is the impact of extreme social and physical distancing measures on the post-lockdown dynamics. Increasing the strength $\alpha$ of the lockdown, not only corresponds to delaying the time at which it is lifted, but it also induces a stronger re-start of the epidemic in the post-lockdown (see APPENDIX for the description of the effects of varying lockdown strength), triggering a new lockdown. Such scenario would obviously be unsustainable, in terms of social and economic costs. 

An additional counter-intuitive mechanism must be considered. Since an attenuation of $\alpha$ corresponds to an effective reproduction number $R_0^{\text{eff}}=\alpha R_0$, at the critical value $\alpha_{\text{crit}}=1/R_0$ the epidemic neither grows nor decreases\footnote{To be precise, the decrease becomes sub-exponential, thus taking a practically infinite time when the size of the population is large}. Thus, after $t_{\text{Lock}}$ the system stays stationary until the lockdown is released at $t_{\text{Unlock}}$; at this point, the epidemic starts growing again as it was before the lockdown. In general, if $\alpha < \alpha_{\text{crit}}$, the system looks to ameliorate (infected, hospitalized, all the infective compartments go down) but as soon as the lockdown is lifted, the epidemic starts again to reach its full strength (see SI). Nevertheless, our estimate $\alpha\sim 0.5 > \alpha_{\text{crit}} \sim 0.3$ for the Italian lockdown gives us hope that, perhaps, it will not be necessary to follow a repeated seek-and-release strategy in the post-lockdown phase. On the other hand, if it can be attained a lockdown strength $\alpha \sim \alpha_{\text{crit}}$ without disrupting the economy, the epidemic could be contained until the creation, production and distribution of a vaccine.

\section{Regional Scenarios}

\begin{figure}
\begin{centering}
\includegraphics[width=0.70\columnwidth]{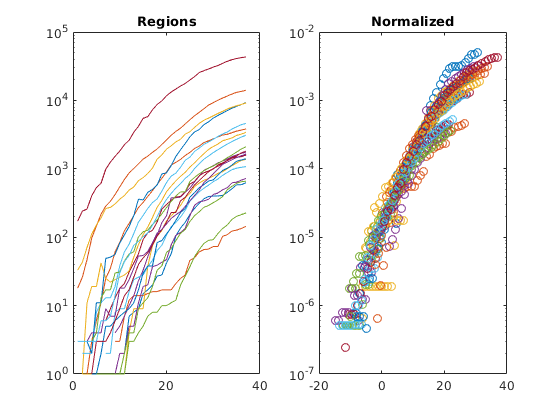} 
\includegraphics[width=0.70\columnwidth]{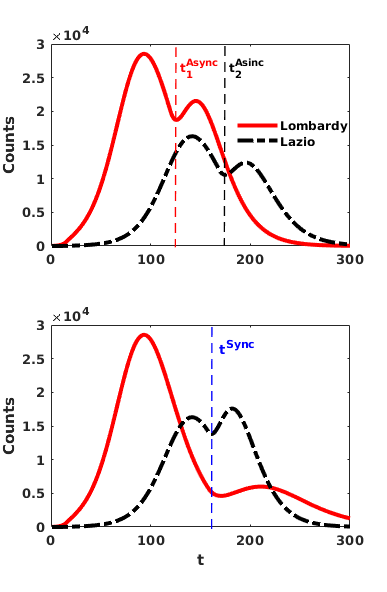} 
\par\end{centering}
\caption{Upper panels: analysis of time delays among the start of epidemics in different regions (see Tab. \ref{tab:Delays}). %Lower panel: effects of the time delays and of different lockdown-end scenarios.
Lower panels: sketch of an \textit{Async}(hronous) exit strategy (i.e. each region lifts the lockdown following its own policy) respect to a \textit{Sync}(hronous) exit strategy (i.e. the lockdown lift follows the same policy, but applied to a nation wide scale). In particular, $t^{Sync}$ corresponds to lifting the lockdown in all the region after the peak has fallen by $30\%$ , while $t^{Async}_i$ corresponds to lifting the lockdown in the $i^{th}$ region after the peak \emph{of such region} has fallen by $30\%$.
}
\label{fig:RegionalScenarios}
\end{figure}

Starting with the first confirmed cases in Lombardy on 21 February, by the beginning of March the Covid-19 outbreak had already spread to all italian regions.
While the delay in the beginning of the infection is accounted for by the different mobility interaction between regions, once the epidemic has started in a given area, the intake of external infected people becomes quickly irrelevant (see APPENDIX for the description of a metaregional model and its behavior). As a consequence, the growth curves of the epidemic variables tend to converge to the same shape (see APPENDIX about using normalised data). In fact, regional info graphics released by the Italian National Healthcare Institute (ISS) \cite{iss_epicentro} show that regional diffusion curves have a similar shape and different starting dates (see Fig.~\ref{fig:RegionalScenarios}).
%More specifically, the earlier to start is Lombardy, the region that up to now has given the highest toll to the Italian epidemics.
This observation can be justified as follows: Italian regions are independent administrative entities, and most of the population tend to work inside the resident region \cite{istat_mobility}. Hence, epidemics propagate from region to region via the fewer inter-regional exchanges (Incidentally, Lombardy is the Italian region, which is most involved in international trade connections \cite{istat_estero}, being the natural candidate for the initial outbreak of the epidemic). More practically, we estimate the delays by minimizing the distance among the observed curves (see APPENDIX for the details of the algorithm); results are reported in Tab.~\ref{tab:Delays}. Notice that, assuming Lombardy has been the first region (i.e. delay=0), the resulting regional delays are mostly correlated to geographical distances.

\begin{table}
\centering
\caption{Regional delays (in days)}
\begin{tabular}{|l c || l c|}
\hline 
Lombardia & 0.0 &  	Molise & 10.6 \\ 
Emilia Romagna & 3.1  & 	Umbria & 11.8 \\ 
Marche & 4.3 & 	Abruzzo & 13.1 \\ 
Veneto & 5.7 & 	Lazio & 14.5 \\ 
Valle d'Aosta & 6.4 & 	Campania & 15.0 \\ 
P.A. Trento & 6.6 & 	Puglia & 15.7 \\ 
P.A. Bolzano & 8.0 & 	Sardegna & 16.2 \\ 
Liguria & 8.1 &	Sicilia & 16.6 \\ 
Friuli Venezia Giulia & 8.9 &  	Calabria & 17.2 \\ 
Piemonte & 9.0 & 	Basilicata & 19.2 \\ 
Toscana & 10.4 & &  \\
\hline
\end{tabular}
\label{tab:Delays}
\end{table}

We assume that the Covid-19 outbreak spreads independently in each region; as argued before, such an approximation is reasonable after the epidemic has started and is even more accurate under lockdown conditions. Hence, we apply the parameters for the entire country to regional cases\footnote{Again, we are exploring qualitative scenarios and we do not aim to predict the real evolution of the epidemics: in fact, Italian regions are different for social contact habits, mobility, organization and capacity of health care provision, as well for factors that affect the medical parameters, like comorbidities, social conditions or pollution levels.}, where now the maximum number of individuals $N_i$ is the population of the $i^{\text{th}}$ region\footnote{https://www.istat.it/it/popolazione-e-famiglie?dati}. Then, by summing up all the $S_i,\ldots ,R_i$, respectively, we obtain a proxy for the global evolution of Covid-19 epidemic throughout Italy. To evaluate the effect of heterogeneity in time delays, we compare the number of daily cases $O^{\text{Delay}}=\sum O_i^{\text{Delay}}$ (obtained by taking into account the regional delays $t_i$ as reported in Tab.~\ref{tab:Delays}) with the number of daily cases $O^{0}=\sum O_i^{0}$ we would observe by considering the epidemics starts at the same time $t_0$ in all regions. As expected, heterogeneity flattens the curve and shifts its maximum later in time. This is a first source of errors when fitting an heterogeneous dynamics with a global model. 

We do take into account regional delays and consider two possible exit strategies: in the first, that we call the \textit{Async}hronous scenario, each region $i$ lifts the lockdown at the time $t^{\text{Unlock}}_i$ when the peak of $O^{\text{Delay}}_i$ decreases by $30\%$; in the second, that we call the \textit{Sync}hronous scenario, each region $i$ lifts the lockdown at the same time $t^{\text{Unlock}}$, i.e., when the global peak of $O^{\text{Delay}}$ decreases by $30\%$. The choice of $30\%$ is arbitrary, similar results would hold for choices of values near the peak; it tries to be a sketch of a situation where, due to economic pressure, lockdown is lifted as soon as possible.

Notice that, once an outbreak has started, the epidemic dynamics in a given region $i$ is essentially uncorrelated with the epidemic spreading of any other region $j\neq i$. Therefore, it could be safe and appropriate to decide the lockdown lifting time on a regional basis, instead of lifting restrictions throughout Italy at the same time. Indeed, it could be unreasonable to keep locked the regions where the epidemic started earlier; on the contrary, regions where the epidemic began with some delay could experience a strong rebound when subjected to a premature lockdown lifting. In Fig.~\ref{fig:RegionalScenarios} we show the effects of lifting the lockdown at both regional (\textit{Async}) and national (\textit{Sync}) level in Lazio and Lombardy.
Since not only epidemics, but also the ruin of an economy is a non-linear process, the \textit{Sync} scenario can turn out to be even more disruptive than the epidemic itself (see also Fig.~\ref{fig:Scenario1}). Notice that analogous arguments hold - mutatis mutandis - also for the world/countries scenario.

\section{The role of Age}

As we have already observed in the previous Section, heterogeneity strongly impacts on the results within the model \cite{diekmann1990heterogeneous}. Since the transmission coefficient is proportional to the contact rate between individuals, the rates of social mixing between different age classes represent a well known important source of heterogeneity. This information can estimated either through large-scales surveys \cite{mossong2008POLYMOD} or through virtual populations modeling \cite{fumanelli2012inferring}. While the POLYMOD \cite{mossong2008POLYMOD} matrices have been extensively used to estimate the cost-effectiveness of vaccination for different age-classes during the 2009 H1N1 pandemic \cite{medlock2009optimizing,baguelin2010vaccination}, here they are used to support the design of a broad class of exit strategies. Hence, to account for age classes, we extend our model by rewriting the transmission coefficient as $\beta C$ (see APPENDIX for a full description of the extended model), where $\beta$ is the transmission probability of the infection, and $C$ is the sociological matrix describing the contact patterns typical of a given country. For lack of further information, we assume $\beta$ constant among age classes and $C$ as in \cite{mossong2008POLYMOD}. To simplify the analysis, we gather POLYMOD age groups into three classes: $Y$oung ($00-19$), $M$iddle ($20-69$) and $E$lderly ($70+$) (see Tab~\ref{tab:contact_matrix}). Such an aggregation puts together the most ``contactful'' classes ($00-19$), the classes with the highest mortality risk ($70+$) \cite{iss_epicentro}, and a good approximation of the active population ($20-69$).

\begin{table}
    \centering
    \begin{tabular}{|c||c|c|c|}
    \hline
      &   Y    &   M    &   E  \\
    \hline
    \hline 
    Y &   2.35 &   0.44 &   0.67\\
\hline
    M &   0.47 &   0.59 &   0.50\\
\hline
    E &   0.50 &   0.55 &   0.80\\
    \hline
    \end{tabular}
    \caption{POLYMOD matrix aggregated for three age classes: $Y$oung ($00-19$), $M$iddle ($20-69$) and $E$lderly ($70+$). }
    \label{tab:contact_matrix}
\end{table}

\begin{figure}
\begin{centering}
\includegraphics[width=0.7\columnwidth]{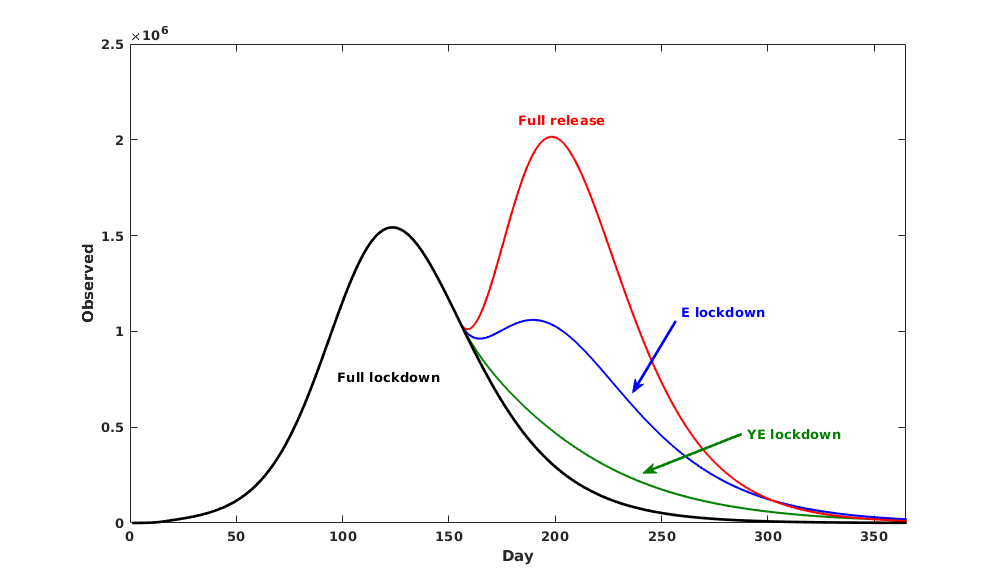}
\par\end{centering}
\caption{Comparison of the scenarios where the lockdown is relaxed only for a particular age class with respect to a full release policy. Strategies: YE = quarantine young and elderly, E = quarantine elderly. Notice that we have purposefully left the M class fully unrestrained, in order to show how maintaining a partial, age-based lockdown could deeply change the effectiveness of the exit strategy.}
\label{diffQ}
\end{figure}

Fig.~\ref{diffQ} shows how the percentage of people with visible symptoms ($O$) varies once the age class heterogeneity is considered in the model. Differently from fig. \ref{fig:Scenario1}, fully lifting the lockdown results in a conspicuous rebound of the epidemics, that reaches values even more severe than the pre-exit peak. Thus, models, which do not explicitly consider this source of heterogeneity could severely mis-forecast the post-exit dynamics. On the other hand, the introduction of the age structure in the model allows to orient the design of exit strategies based on age-targeted policies, as a way to dampen a possible upturn of contagion. Specifically, social/physical distance measures applied to the elderly may contribute to contain the impact of a renewed upward phase, while relaxing restrictions to the working age class (20-69) would not impair the smoothing of contagion propagation in the post-lockdown phase. Again, a disclaimer, it is important to emphasize that we are referring to simple mock-up strategies, which correspond to worst-case scenarios: in real life, community measures and physical distancing, infection prevention and control, personal hygiene habits, face mask usage, etc. will be decisive in contributing to the dampening of the epidemics \cite{ECDC8apr2020report}.  

\section{Conclusions}

In this paper we propose and test a general framework to study the Covid-19 contagion through a compartmental model, with a focus on geographical groups and age classes. Our framework shows that the promptness of lockdown measures has a main effect on the timing of the contagion. Strict social distancing policies reduce the severity of the epidemics during the lockdown period, but full recover of the contagion can occur once such measures are relaxed. As a consequence, a mix of specific mitigation strategies must be prepared during the lockdown and implemented thereafter. In order to understand the relative potential impact of different broad strategies, we focus on two broad decomposition criteria within the model, that is geographical mobility and social interactions between age classs. Our results are driven by the sparsity of the underlying contact matrices, which we measure. First, we show how local dynamics at regional level can be hidden when observing the aggregate national system. Regional heterogeneity tends to lower and widen the curve of the contagion, contributing to a shift forward in time for the peak at the aggregate level. Moreover, our analysis of mobility data shows that, due to the sparsity of interconnections across regions, contagion develops independently within each region once the epidemic has started. This, in turn, contributes to account for the delays observed in the alignment of the contagion curves across different geographical areas. The independence of regional dynamics is important, since it can justify the adoption of a mix between general mitigation strategies and solutions which are specific to individual regions or to clusters of regions. Interestingly enough, the generality of our model makes this result relevant also to frame the relative impact of cross country mobility flows. Finally, we investigate the structure of social contacts across different settings and we quantify the relative importance of interactions between age classes in the spreading of contagion. We show that the young (0-19) and the old (70+) are the most intensively interacting classes. As a consequence, mitigation strategies specific to these two classs can produce a significant impact on diffusion rates in the post-lockdown phase. In fact, our results show the importance of designing physical distancing measures specific to the elderly and, in addition, they can orient decisions on limitations to social contacts for the young. Overall, our results provide guidance on how to relax some of the restrictions to mobility for the active population (20-69), while smoothing and lessening the propagation of contagion in the post-lockdown phase.

Although our study is tuned on the Italian Covid-19 contagion, our modeling approach is general enough to help us understand the role of relevant dimensions, beside the medical and pharmaceutical ones, in identifying the relative importance of different strategies introduced to contain the epidemics and to mitigate its effects. Our framework can contribute to mitigate the stringency of the trade-off between health and economic outcomes. In particular, we show how the timeline of post-lockdown measures should take into account some fundamental compartmental aspects, such as geographical factors and interactions between different age classes. This feature is general, and it can orient the analysis towards fine grained simulations on the impact of specific precautionary interventions, which enforce social distancing while containing the overall burden on the economy and on society. 

%\subsubsection*{Acknowledgements}
%
%A.S., A.F, F.P,  E.B., M.C. and W.Q. acknowledge the support from CNR P0000326 project AMOFI (Analysis and Models OF social medIa) and CNR-PNR National Project DFM.AD004.027 `` Crisis-Lab''.

%\clearpage
%\newpage
\bibliographystyle{unsrt}
\bibliography{SceCov}

\clearpage
\newpage
\section{APPENDIX \label{sec:APPENDIX}}

In Sec. \ref{SI:model} we describe the $SIOR$ model used in the paper, stressing the general problems related to fitting real data with compartmental models. In Sec. \ref{SI:locktimestrength} we discuss the effects of varying the starting date and the strength of the lockdown measures. The algorithm used for finding delays among growth curves is described in Sec. \ref{SI:delays}, whereas the reasons why classic growth curves could have the same shape when normalised is discussed in Sec. \ref{SI:normalized}. Moreover, Sec. \ref{SI:metaregional} explicates why from an extension of a simple compartmental model to a regional metapopulation model with a very low mobility among regions it is to be expected that regions show similar dynamics like in Sec. \ref{SI:normalized} but shifted in time. Finally, Sec. \ref{SI:socialmix} shows the extension of a simple compartmental model to consider social mixing among different age classes.

\subsection{Basic model\label{SI:model}}
\begin{figure}
\centering
\includegraphics[width=0.7\columnwidth]{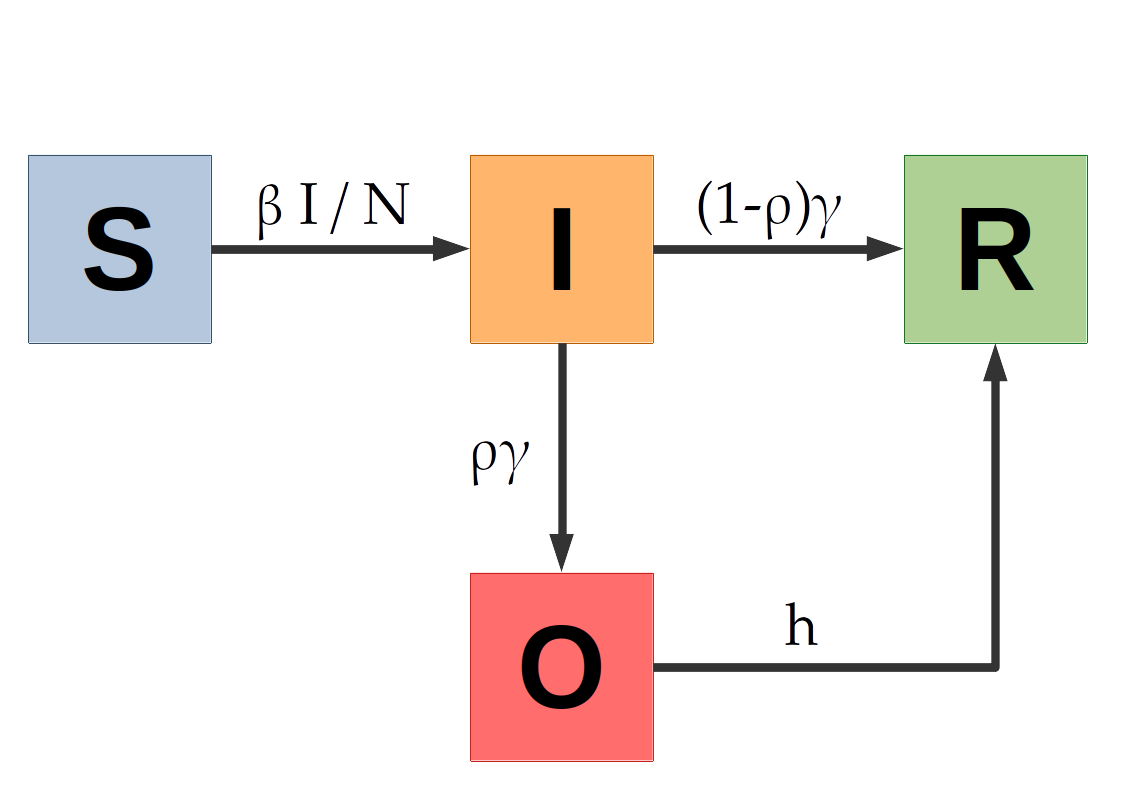}
\caption{The $SIOR$ compartmental model: workflow of the epidemic process. A $S$(usceptible) individual becomes $I$(nfectious) when meeting an infected person. An $I$(nfectious) either become $O$(bserved), with symptoms acute enough to be detected from the national health-care system, or is $R$(emoved) from the infection cycle by having recovered. An $O$(bserved) individual can also be $R$(emoved) from the infection cycle having become immune.
The parameter $\beta$ defines the rate at which a susceptible becomes infectious, $\gamma$ represents the rate at which infectious either become observable or recover, $\rho$ is the fraction of infectious that become observed from the national health-care system and $h$ is the rate at which observed individuals are removed from the infection cycle.}
\label{fig:SIOR}
\end{figure}

Our model belongs to the classic family of compartmental models \cite{Bailey1975book}. As the most renewed $SIR$ and $SEIR$ model (and their variations), it models the infection rate to be proportional to the number of individuals in a $S$(usceptible) compartment (i.e., the ones that have never been infected) times the probability of meeting infected persons (modelled as the fraction $I/N$ of individuals in the $I$(nfectious) compartment respect to the population size $N$). The other essential rate is represented by individuals that are $R$emoved (either because recovered and not more susceptible, or because deceased) from the $I$ class; again, such rate is proportional to the number of individuals in $I$. To have the possibility of adjusting our model's parameter with the observed data, we introduce another class $O$ of ``observable'' people, i.e. people with symptoms strong enough to be detected by the national healthcare system. A graphical sketch of the model is presented in fig.\ref{fig:SIOR}.    

The $SIOR$ model is described by the following set of ordinary differential equations:
\begin{equation}
\begin{split}
\partial_t S &= - \beta S\,\frac{I}{N} \\
\partial_t I &= \beta S\,\frac{I}{N} - \gamma I \\
\partial_t O &= \rho \gamma I - h O \\
\partial_t R &= (1-\rho) \gamma I + h O
\end{split}
\label{eq:SIOR}
\end{equation}
where $N=S+I+O+R$ is the total number of individuals in a population, the transmission coefficient $\beta$ is the rate at which a susceptible becomes infected upon meeting an infected individual, $\gamma$ is the rate at which an infected either becomes observable or is removed from the infection cycle. The extra parameters of the model are $\rho$, the fraction of infected that become observed from the national health-care system, and $h$, the rate at which observed individuals are removed from the infection cycle. Notice that we consider that $O$(bserved) individuals not infecting others, being in a strict quarantine.
 
As for the $SIR$ model, the basic reproduction number can be calculated as $R_0=\beta/\gamma$ and the stationary state can be estimate as follows. Let us consider $X=O+R$, it holds $\partial_X S =-R_0 S$ and $S(t\to\infty)=Ne^{-R_0 X(t\to\infty)}$. Then, since $O(t\to\infty)=I(t\to\infty)=0$, it follows that $R(t\to\infty)=N-S(t\to\infty)$ and we recover the same solution of the $SIR$ model: $S(t\to\infty)=Ne^{-R_0\,[N- S(\to\infty)]}$. 
%For $R_0 \in [2.5,4.5]$, we have that the final fraction of uninfected population varies between $10\%$ and $1\%$.

\subsubsection{Initial parameters estimation\label{SI:initialparams}}
\begin{figure}
\begin{centering}
\includegraphics[width=0.90\columnwidth]{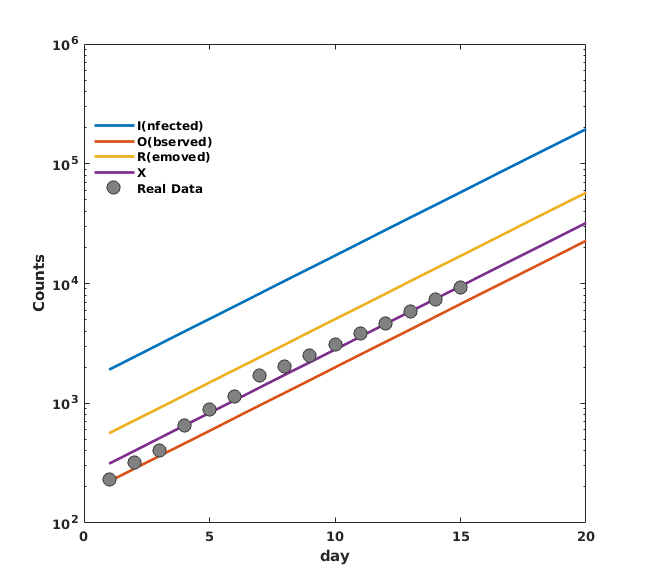}
\par\end{centering}
\caption{In the initial stage, most of the quantities experience an exponential growth with the same exponent; hence, it would be possibly to ``successfully'' fit the wrong variables.  Figure shows the pre-lockdown growth of the number of $I$(nfected), $O$(bserved), $R$(emoved) individuals in our model (\protect{\ref{eq:SIOR}}). Full circles represent the experimental counts of confirmed Covid-19 cases in Italy; $X$ is the cumulative variable we use to fit the experimental data.}
\label{fig:Early}
\end{figure}

In the early phases of the epidemic, the observed quantities follow an approximately exponential growth $Y^{\text{Obs}}\sim Y_0e^{gt}$, as expected in most epidemic models. To understand what happens in our model, we notice that for $I/S\ll 1$ we can linearize System~\ref{eq:SIOR} resulting in $I\sim I_0 e^{(\beta-\gamma)t}$ and $O\sim \rho\gamma I$. Thus, minimizing the difference between $O$ and $Y^{\text{Obs}}$ in the early period would yield estimates for $\beta,\gamma$ such that $\beta-\gamma\sim g$, and $R_0\sim 1+g/\gamma$ would increase linearly with the characteristic time $\tau_I=\gamma^{-1}$ for exiting the infectious phase. Notice that most of the compartmental models based on a set of ordinary differential equations show an initial exponential growth phase with the same exponent (see Fig.~\ref{fig:Early}); hence, in the early stage of the epidemic it is possible to successfully fit the ``wrong'' variables.

\subsection{Effects of lockdown time and strength\label{SI:locktimestrength}}

\begin{figure}
\begin{centering}
\includegraphics[width=0.4256\columnwidth]{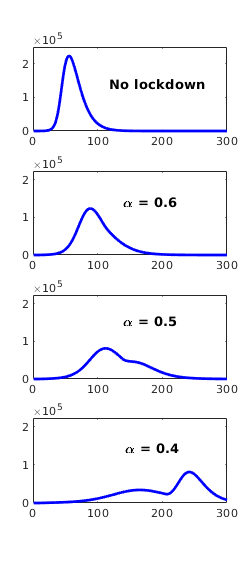} \includegraphics[width=0.4\columnwidth]{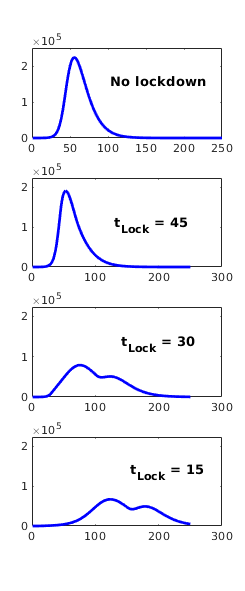}
\par\end{centering}
\caption{Left panel: variation of the behavior of the model by varying the lockdown strength $\alpha$. Lockdown starts at $t_{\text{Lock}}=15$ and is fully lifted when the peak has fallen by $30\%$. Right panel: variation of the behavior of the model by delaying the lockdown time $t_{\text{Lock}}$. Lockdown strength is fixed at $\alpha=0.5$ and is fully lifted when the peak has fallen by $30\%$.}
\label{lockstrengthtime1}
\end{figure}

By increasing the strength $\alpha$ of the lockdown (where $\alpha$ is the ratio between the trasmission $\beta$ after and before the lockdown) the epidemic peak is pushed forward, but the height of it is lower. On the other hand, the epidemic delaying implies that lifting the lockdown would bring back the infection. In the left panel of Fig.~\ref{lockstrengthtime1}, we show what happens by lifting the lockdown when the peak is fallen by $30\%$: stronger lockdowns induce a stronger reprise of the epidemic. An analogous effect can be observed by varying the lockdown time: anticipating the lockdown ameliorates the peak by decreasing its height, but shifts it to later time and retards the end of the epidemic. 

\begin{figure}
\begin{centering}
\includegraphics[width=1\columnwidth]{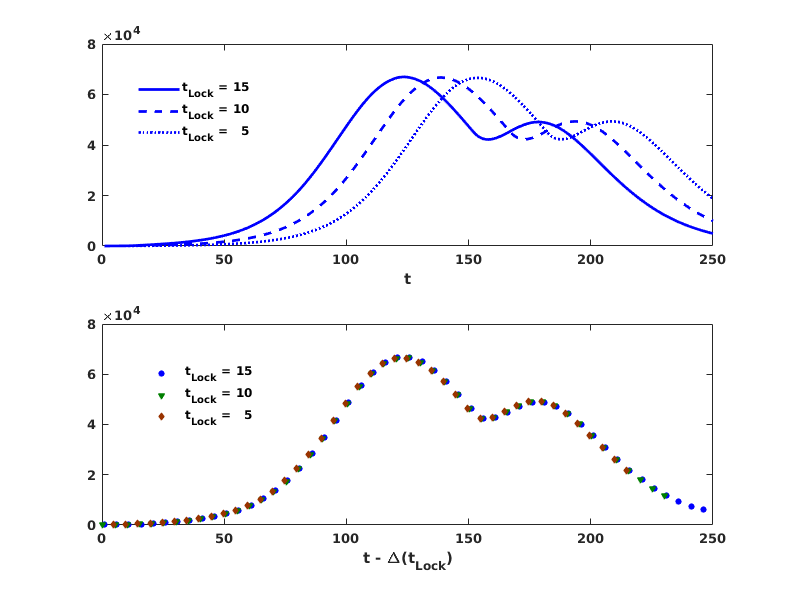}
\par\end{centering}
\caption{Upper panel: variation of the behavior of the model by anticipating the lockdown time. Notice that anticipating the lockdown leaves unchanged the behaviour of the epidemics, just shifting all the times of an amount proportional to how much the lockdown is anticipated.  Lockdown strength is fixed at $\alpha=0.5$ and is fully lifted when the peak has fallen by $30\%$. Lower panel: by applying the Eq.~\protect{\ref{eq:lockshift}}, we show how the curves in the upper panel collapse on each other.}
\label{figlockshift}
\end{figure}

Contrary to what could be naively expected, an early imposition of the lockdown does not ameliorate the epidemics: in fact, anticipating too much the lockdown just shifts the timing of the epidemics, leaving its evolution unchanged (see Fig.~\ref{figlockshift}). This is to be expected every time extreme measures of social distancing are applied in the very early, exponentially growing, stages. In fact, let us consider two countries $A$ and $B$ that have the same population, the same contact matrix, and the same number of infected persons. If $A$ and $B$ decide to put a lockdown of strength $\alpha$ at time $t_A$ and $t_B$, respectively, at time $t$ any quantity $y$ of the model would have grown as $y_A(t)\sim y^0\,e^{R_0 t_A} \,e^{\alpha R_0(t-t_A)}$ and as $y_B(t)\sim y^0 \,e^{R_0 t_B} \,e^{\alpha R_0(t-tB)}$. If there exists a $t'$ such that $y_A(t)=y_B(t')$, the epidemics in $A$ and in $B$ will proceed in parallel (even in the non-linear phase) with a delay $t'-t$. Therefore, if the epidemic dynamics of $A$ and $B$ are still well approximated by exponential distributions at times $<\max\{t,t'\}$, then $t'-t\propto -(t_A-t_B)$, i.e, the country that has started the lockdown before will experience the same epidemic of the other country, just delayed in time. In particular, for identical initial conditions, we have that: 
\begin{equation}
t-t'=-\frac{1+\alpha}{\alpha}(t_A-t_{B}) 	\label{eq:lockshift}
\end{equation}
as long as all the times are before the initial exponential regime ends. Such an estimate can be very useful for countries where the epidemics has not started yet. Indeed, calibrating on one own normalized growth curve the time of the lockdown and its strength would give an idea of how long one can delay the full start of the epidemic dynamics.

\begin{figure}
\begin{centering}
\includegraphics[width=1\columnwidth]{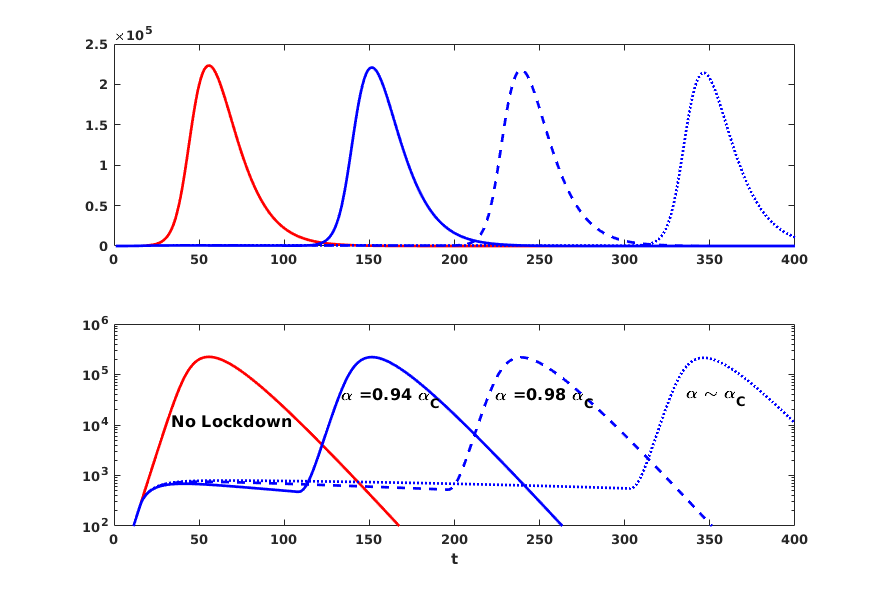}
\par\end{centering}
\caption{Upper panel: variation of the behavior of the model for lockdown strengths $\alpha<\alpha_{\text{crit}}=1/R_0$. Notice that  the height of the peaks after the lockdown is released is almost unchanged if compared with the no lockdown scenario.  Lockdown time is fixed at $t_{\text{Lock}}=15$ and is fully lifted when the peak has fallen by $30\%$. Lower panel: for better clarity, the plot is also reported in log-linear scale.}
\label{fig:lockshift}
\end{figure}

Finally, we notice that to each lockdown strength $\alpha$ corresponds an effective reproduction number $R_0^{\text{eff}}=\alpha R_0$; hence, for $\alpha \sim \alpha_{\text{crit}}=1/R_0$, the epidemics is expected to stay in a quiescent state where it does not either grow or decay sensibly. On the other hand, for $\alpha<\alpha_{\text{crit}}$ the epidemics decreases; nevertheless, since this happens before a sufficient number of recovered individuals has built up herd-immunization, the height of the peaks after the lockdown lifting are almost unchanged if compared with the no lockdown scenario. Again, a ``too good'' intervention risks to postpone the problem without attenuating it. Notice that, if one applies lockdowns with $\alpha<\alpha_{\text{crit}}$, it could be necessary to switch back and forth to lockdown to avoid the peak go beyond the capacity of a national healthcare system (see Fig.~\ref{fig:lockshift}). 

\subsection{Estimation of the experimental time delays\label{SI:delays}}
We first normalize the observed data by dividing the number of non-zero observations in a region for the population of the region. Let $y_i$ be the normalized observations for the $i^{\text{th}}$ region. 
For each pair of regions $i,j$, we define the variation interval $\Delta_{ij}=[\min_{ij},\max_{ij}]$ that contains the maximum number of points of both $y_i$ and $y_j$, i.e. $\min_{ij}=\max\{\min(y_i),\min(y_j)\}$ and $\max_{ij}=\min\{\max(y_i),\max(y_j)\}$. 
The delay $t_{ij}$ between the epidemics start in $i$ and $j$, respectively, is calculated by minimizing the square norm of $\lVert(\Delta_{ij} \cap y_i(t))\setminus(\Delta_{ij} \cap y_j(t-t_{ij})\rVert$, where $\Delta_{ij} \cap y$ denotes the values of $y$ falling in the interval $\Delta_{ij}$. 
Denoting with $T_i$ the times corresponding to the observation in $\Delta_{ij} \cap y_i$, it is easy to verify that $t_{ij}=\langle T_i\rangle-\langle T_j\rangle$, where $\langle T\rangle$ is the average value of the times contained in $T$.

\section{Equivalence of normalized curves\label{SI:normalized} }

Eq.~\ref{eq:SIOR} referred to region $k$ becomes:
\begin{equation}
\begin{split}
\partial_t S_k &= - \beta S_k\,I_k/N_k \\
\partial_t I_k &= \beta S_k\,I_k/N_k - \gamma I_k \\
\partial_t O_k &= \rho \gamma I_k - h O_k \\
\partial_t R_k &= (1-\rho) \gamma I_k + h O_k
\end{split}
\label{eq:SIORegion}
\end{equation}
where $N_k$ is the population of the region. By rewriting Eq.~\ref{eq:SIORegion} in terms of normalized quantities $s_k=S_k/N_k,\ldots s_k=S_k/N_k$, we obtain the same equation for all the regions:
\begin{equation}
\begin{split}
\partial_t s &= - \beta s\,i \\
\partial_t i &= \beta s\,i - \gamma i \\
\partial_t o &= \rho \gamma i - h o \\
\partial_t r &= (1-\rho) \gamma i + h o
\end{split}
\label{eq:sior}
\end{equation}
Hence, for similar initial conditions, by normalizing the experimental observations by the population, one should obtain similar time behaviors.

\subsection{Regional metapopulation model\label{SI:metaregional} }
Let us assume that we know the fraction $T_{kl}$ of people commuting from region $k$ to region $l$, Eq.~\ref{eq:sior}  becomes:
\begin{equation}
\begin{split}
\partial_t s_k &= - \beta s_k\,\sum_l T_{kl}\, i_l \\
\partial_t i_k &= \beta s_k\,\sum_l T_{kl}\,i_l - \gamma\,i_k \\
\partial_t o_k &= \rho \gamma\, i_k - h\,o_k \\
\partial_t r_k &= (1-\rho) \gamma\, i_k + h\,o_k
\end{split}
\label{eq:metasior}
\end{equation}
From mobility data, we know that $\epsilon_k = \sum_{l\neq k}T_{kl} / T_{kk} \ll 1 $ and $T_{kk}\sim 1$; in particular, from Facebook mobility data we can estimate $\langle\epsilon_k\rangle\sim 10^{-3}$.
If all the neighbors of a given region $k$ are fully infected (i.e. $i_l=1\,\,\forall\,{l\neq k}$) and $i_k(t_0)=0$, then the variation of $i_k$ can be approximated as $\partial_t i_k \sim \epsilon_k +(\beta-\gamma) \,i_k $. Namely, as soon as $i_k>\epsilon_k$, $i_k$ will grow exponentially according to $\partial_t i_k  \sim (\beta-\gamma) \,i_k$ and $\epsilon_k$ will become irrelevant; that is to say, the dynamics of the regions will decouple. On the other hand, if epidemic is decaying everywhere, then $i_l\ll 1\,\,\forall\,{l\neq k}$; thus $\sum_{l\neq k} T_{kl}\, i_l \ll \epsilon_k$ and equation again decouple, having each region followed Eq.~\ref{eq:sior} separately. In Tab.~\ref{tab:MobVsDelay} we confront regions ordered by simulating an hypothetical epidemics starting from Lombardy and propagating with Eq.~\ref{eq:metasior}, with regions ordered by the estimated delays obtained by applied the algorithm of sec. \ref{SI:delays}. It is reasonable to assume that inter-regional mobility has had a role in the regional delay structure; however, many other factors come to play in the long range propagation of epidemics: as an example, both airline transportation network \cite{hufnagel2004forecast,guimera2005worldwide} and individual work commutes \cite{hall2007comparison,viboud2006synchrony} have played important roles in understanding the spread of infectious diseases.

\begin{table}
    \centering
\scalebox{0.72}{    \begin{tabular}{|c|c|}
    \hline
        Mobility Matrix & Experimental Delays \\
    \hline
        Lombardia & Lombardia \\
        Emilia Romagna & Emilia Romagna \\
        Piemonte & Marche \\
        Veneto & Veneto \\
        Valle d'Aosta & Valle d'Aosta \\
        Trentino Alto Adige & Liguria \\
        Lazio & Friuli Venezia Giulia \\
        Liguria & Piemonte \\
        Toscana & Trentino Alto Adige \\
        Campania & Toscana \\
        Marche & Molise \\
        Friuli Venezia Giulia & Umbria \\
        Abruzzo & Abruzzo \\
        Umbria & Lazio \\
        Sardegna & Campania \\
        Sicilia & Puglia \\
        Molise & Sardegna \\
        Basilicata & Sicilia \\
        Puglia & Calabria \\
        Calabria & Basilicata \\
    \hline
\end{tabular}}
    \caption{Region ordered by simulations using the mobility matrix (left column) and by the delays obtained by rescaling experimental data (right column).}
    \label{tab:MobVsDelay}
\end{table}

\subsection{Social mixing}
\label{SI:socialmix}
To take account for social mixing,  we rewrite the transmission coefficient as the product of a transmission probability $\beta$ times a contact matrix $C$ whose element $C_{ab}$ measure the average number of (physical) daily contacts among an individual in class age $a$ and an individual in class age $b$. Notice that the probability that a susceptible in class $a$ has a contact with an infected in class $b$ is the product of the contact rate $C_{ab}$ times the probability $I_b/N_B$ that individual in class $b$ is infected. Hence, denoting with $S^a,\ldots,R^a$ the number of $S$(usceptibles),$\ldots$,$R$(emoved) individuals in class age $a$, we can rewrite Eq.~\ref{eq:SIOR} as: 
\begin{equation}
\begin{split}
\partial_t S^a &= - \beta S^a\sum_b  C_{ab} \frac{ I^b}{N_b} \\
\partial_t I^a &= \beta S^a\sum_b  C_{ab} \frac{ I^b}{N_b}  - \gamma I^a \\
\partial_t O^a &= \rho \gamma I^a - h O^a \\
\partial_t R^a &= (1-\rho) \gamma I^a + h O^a
\end{split}
\label{eq:SIORage}
\end{equation}
Although the form of Eq.~\ref{eq:SIORage} is similar to Eq.~\ref{eq:SIORegion}, here it is not possible to consider separate evolutions for the different age classes since, differently than the inter-regional mobility matrix $T$, the off diagonal elements of the social matrix $C_{a,b}, a\neq b$, measure the interaction among different age classes and are of the same magnitude of the diagonal elements $C_{aa}$ measuring the interaction among individuals of the same age class.

Notice that Eq.~\ref{eq:SIORage} can be summed up, and the resulting equation can be obtained by substituting $\beta\to \beta C^{\text{eff}}$ in Eq.~\ref{eq:SIOR}, where $C^{\text{eff}}= \frac{ \sum_{ab} C_{ab} S^a I^b/N_b}{SI/N}$ is the average contact value among infected and susceptible individuals of all age classes.

\end{document}